\documentclass[aps,prc,twocolumn,superscriptaddress,showpacs,floatfix]{revtex4}
\usepackage[dvips]{graphicx}
\usepackage{amsmath,amssymb}

\begin{document}
\title{ Parametrizations of triaxial deformation
       and $E2$ transitions of the wobbling band }

\author{Yoshifumi R. Shimizu}
\affiliation{Department of Physics, Graduate School of Sciences,
Kyushu University, Fukuoka 812-8581, Japan}

\author{Takuya Shoji}
\affiliation{Department of Physics, Graduate School of Sciences,
Kyushu University, Fukuoka 812-8581, Japan}

\author{Masayuki Matsuzaki}
\affiliation{Department of Physics, Fukuoka University of Education,
             Munakata, Fukuoka 811-4192, Japan}

\date{\today}

\begin{abstract}
By the very definition the triaxial deformation parameter $\gamma$
is related to the expectation values of the $K=0$ and $K=2$ components
of the intrinsic quadrupole tensor operator.
On the other hand, using the same symbol "$\gamma$",
various different parametrizations of triaxial deformation
have been employed,
which are suitable for various types of the mean-field potentials.
It is pointed out that the values of various "$\gamma$"
are quite different for the same actual triaxial deformation,
especially for the large deformation; for example,
the difference can be almost a factor two for the case of
the triaxial superdeformed bands recently observed in the Hf and Lu nuclei.
In our previous work, we have studied the wobbling band in Lu nuclei
by using the microscopic framework of
the cranked Nilsson mean-field and the random phase approximation,
where the most serious problem is that
the calculated $B(E2)$ value is about factor two smaller.
It is shown that
the origin of this underestimation can be mainly attributed to
the small triaxial deformation;
if is used the same triaxial deformation as
in the analysis of the particle-rotor model,
the calculated $B(E2)$ increases and gives correct magnitude
compared with the experimental data.

\end{abstract}

\maketitle

\section{Introduction}
\label{sec:intro}

The triaxial deformation of atomic nuclei has been one of
the long standing issues in nuclear structure physics.
If not going back to the Davydov-Phillipov model~\cite{DF58},
the triaxial rotor model is very useful to study
the spectra of odd transitional nuclei~\cite{Mey75}.
More recently, the model-independent sum-rule method~\cite{Cl86}
is used to analyse the Coulomb excitations
for the quadrupole collective motions,
where the triple $E2$ matrix elements is related the the triaxiality
parameter $\gamma$.  Many of Coulomb excitation measurements
revealed clearly non-axial deformation~\cite{AP94},
although it was difficult to distinguish
that the non-axiality is of static or of dynamic nature.
At high-spin states, it was expected
that the effect of triaxial deformation
appears more explicitly in the excitation energies
and/or $M1$/$E2$ transition probabilities,
see \cite{HamRev1,HamRev2} and references therein;
for example, the so-called signature
staggering in odd or odd-odd nuclei
was supposed to be a good indicator of triaxiality,
but the result was not so definite as expected.

However, the situation has been changed quite recently:
The nuclear wobbling motion~\cite{BM75} has been identified
in some Lu isotopes,
$^{163}$Lu~\cite{wob163Lu1,wob163Lu2,wob163Lu3,wob163LuTwo},
$^{165}$Lu~\cite{wob165Lu},
$^{167}$Lu~\cite{wob167Lu}, and $^{161}$Lu~\cite{wob161Lu},
and it is now believed that
we are able to study how a triaxially deformed nucleus rotates.
In fact it has been predicted that the strong triaxial deformation
appears in this mass region~\cite{Rag89,Ab90,Bengt},
i.e. the Hf and Lu region,
where the wobbling phonon excitations have been measured.
The triaxial deformation predicted in such nuclei is
the so-called positive $\gamma$ shape in the Lund convention~\cite{NR95},
i.e. nuclei rotate about the shortest axis,
with the Nilsson deformation parameters
$(\epsilon_2,\gamma)\approx (0.43,20^\circ)$,
and the associated rotational sequence is called
the triaxial superdeformed (TSD) band~\cite{SP95}.

In order to pin down how much the TSD nucleus deforms triaxially,
in which the wobbling excitation is measured,
one needs to have recourse to some model.
The standard one is the triaxial rotor model~\cite{DF58},
where the high-spin states should be considered~\cite{BM75}.
In this model, the effect of triaxial deformation appears primarily
in the three different moments of inertia and the two intrinsic
quadrupole moments of nucleus.
The formers are mainly responsible to the excitation energies,
while the latters are directly reflected in the $E2$ transition probabilities.
Compared with the moments of inertia which are sensitive to other factors,
like the pairing correlations, the $E2$ transition probabilities
are more direct and robust quantities, and therefore,
the measurements of $B(E2)$~\cite{wobLuQ,wob163LuQ},
not only that of in-band (intraband) but of out-of-band (interband)
between the yrast TSD band and the excited wobbling band, are crucial
to obtain the information about the triaxial deformation.
The detailed study by using
the particle-rotor model~\cite{wob163Lu1,Ham02,HH03},
where an odd $i_{13/2}$ proton is coupled to the triaxial rotor,
which is suitable for the description of odd Lu TSD bands,
revealed that
the observed ratio of out-of-band to in-band $B(E2)$'s,
$B(E2)_{\rm out}/B(E2)_{\rm in}$, is consistent to the triaxiality
parameter $\gamma\approx 20^\circ$.

In this respect, it is worthwhile mentioning Ref.~\cite{CMZ03},
where it is discussed that the out-of-band $B(E2)$ from the
one-phonon wobbling band is not enough to distinguish
the $\gamma$-soft and $\gamma$-rigid triaxialities in nature.
Further, it is proposed that the measurement of the $B(E2)$
between the odd-spin and even-spin members of
the wobbling excitations is crucial to distinguish the two.
Here even-even nuclei are considered in~\cite{CMZ03}
so that the observed one-phonon wobbling band corresponds to
the odd-spin members for the even-spin yrast TSD band.
In some Lu isotopes the so-called two-phonon wobbling bands
have been observed~\cite{wob163LuTwo,wob165Lu}, which corresponds to
the even-spin members of the wobbling excitations for even-even nuclei.
The $B(E2)$ values from the two-phonon to one-phonon wobbling band
are measured to be about two times the values
from the one-phonon wobbling to yrast band,
which clearly fits the picture of the $\gamma$-rigid model rather than
the $\gamma$-soft model.
Although the negative $\gamma$ shape is assumed in Ref.~\cite{CMZ03},
which is believed to be opposite to
what is measured in Lu isotopes (see below),
it does not affect the main issue; whether
the $\gamma$-soft or $\gamma$-rigid model is superior.
In this way, it is more likely that the triaxial deformation is
of static nature, i.e. the rigid model is more suitable,
and it is meaningful to ask
how much triaxial deformation the observed TSD bands have.

It should, however, be noticed that the conventional
macroscopic rotor model
with irrotational moments of inertia has an essential problem;
the rotor rotates around the intermediate axis,
which corresponds to the negative $\gamma$ shape,
and conflicts to the measured $B(E2)$ ratio.
Therefore, in Refs.~\cite{wob163Lu1,Ham02,HH03},
the largest and intermediate moments of inertia are interchanged
to simulate the positive $\gamma$ shape.
It is well-known that the nuclear moments of inertia are neither
that of the rigid-body nor of the irrotational fluid,
and microscopic models are necessary for their proper description.
Thus, we have investigated the wobbling motion
in the Lu region~\cite{MSMwob1,MSMwob3,MSMwob2}
by employing a microscopic framework, the cranked mean-field
and the random phase approximation (RPA)~\cite{Mar77,JM79,Mar79}.
This approach is suitable to describe the vibrational excitations
in the rapidly rotating nuclei,
see e.g. Refs.~\cite{EMR80,SM83,SM84,KN04,KN06};
and have been used to study possible wobbling excitations
in normal deformed nuclei in our previous works~\cite{Mat90,SM95},
and more recently in Refs.~\cite{AND06,KN07}.

In our approach the residual interaction is chosen consistently
to a given mean-field, and there is no ambiguity of
the force parameters for the RPA calculation.
It has been found~\cite{MSMwob1,MSMwob3,MSMwob2}
that the RPA solutions, which can be nicely interpreted as
the wobbling phonons, do exist in the Lu region for a suitable
range of deformation parameters corresponding to the prediction
of the TSD bands.  The calculated excitation energies is in
a reasonable range in comparison with the experimental data,
though they do not fit the data very precisely.
However, the calculated $B(E2)$ ratios are systematically
smaller by about a factor two to three,
as long as is used the triaxiality parameter
$\gamma\approx 20^\circ$, which is predicted by the Nilsson-Strutinsky
calculations~\cite{SP95,Bengt} and
is used in the analysis of the $B(E2)$ ratio by
the particle-rotor model~\cite{HH03}.
The main issue of the present work is to discuss
why our RPA calculations of the wobbling motion
underpredict the $B(E2)$ ratio.  In the course of discussion,
it is clarified that the various definitions of the triaxiality
parameter $\gamma$ give rather different values
and one has to be very careful when talking about the triaxial
deformation especially for a larger deformation
like in the case of TSD bands.

The paper is organized as follows:
Various existing definitions of the triaxiality parameter $\gamma$
are reviewed, and their values for a given shape are calculated and
compared in \S\ref{sect:tripara}.  Some general discussion on
the relations between various definitions are also given there.
After discussing the difference between the $\gamma$ values
used in the Nilsson-Strutinsky calculations and the particle-rotor model,
in \S\ref{sect:be2ratio},
it is shown that our RPA calculation gives a correct magnitude
of the $B(E2)$ ratios if is used the corresponding triaxial
deformation to the analysis of the rotor model.
\S\ref{sect:sum} is devoted to the summary.
A part of the present work was presented
in some conference reports~\cite{SMM06,SS06}.

\section{Parametrizations of triaxial deformation}
\label{sect:tripara}

The amount of triaxial deformation is usually designated
by the triaxiality parameter $\gamma$,
but there are various definitions for it.
In this section, we discuss the relations between them
and show how large their differences are for a given shape.
It should be mentioned that this problem has been already
discussed in Ref.~\cite{SM84IVD} (Appendix B) for
the volume-conserving anisotropic harmonic oscillator potential.
The present study generalizes its conclusion to more realistic
potentials.

\subsection{Basic definition based on the quadrupole moments}
\label{sect:gamQmom}

One of the most important characteristics of nuclei with
(static) triaxial deformation is the existence of two distinct
intrinsic quadrupole moments.  In this paper we assume
the intrinsic $z$-axis as a quantization axis and the $x$-axis
as a rotation axis, and define
the two moments~\cite{BM75},
\begin{equation}
\left\{\begin{array}{l}
 Q_0\equiv\sqrt{\frac{16\pi}{5}}\langle \hat{Q}_{20} \rangle
 ={\displaystyle
 \int (2z^2-x^2-y^2)\rho(\mbox{\boldmath$r$})d^3\mbox{\boldmath$r$}
 },\\
 Q_2\equiv\sqrt{\frac{16\pi}{5}}\langle \hat{Q}_{22} \rangle
 =\sqrt{\frac{3}{2}} {\displaystyle
 \int (x^2-y^2)\rho(\mbox{\boldmath$r$})d^3\mbox{\boldmath$r$}
 }, \end{array}\right.
\label{eq:Qintr}
\end{equation}
where $\hat{Q}_{2K}\,(K=0,\pm 1,\pm 2)$ are the usual quadrupole operator
in the intrinsic frame of the deformed nucleus,
and $\rho(\mbox{\boldmath$r$})$ is the nucleonic density.
These two moments are directly related to the in-band and
out-of-band $B(E2)$ values of the wobbling band
according to the rotor model~\cite{BM75},
and their measurements in the Lu isotopes~\cite{wobLuQ,wob163LuQ}
uniquely determine these moments.
In place of the two moments, equivalent two quantities,
the magnitude of moments $Q$ and
the triaxiality parameter $\gamma$ are usually used:
\begin{equation}
 Q_0 = Q \cos\gamma,\quad -\sqrt{2}\,Q_2 = Q \sin\gamma.
\label{eq:Qgamma}
\end{equation}
Here we follow the Lund convention~\cite{NR95}
of the sign of the triaxiality parameter $\gamma$,
which is opposite to that of Ref.~\cite{BM75}.
In the following we are mainly concerned with the parameter $\gamma$,
and  consider the in-band and out-of-band $B(E2)$ ratio
of the one-phonon wobbling band, which is independent of the magnitude
of the moment $Q$; see the next section.

The triaxiality parameter defined above reflects the nuclear density
distribution and we call it ``$\gamma(\mbox{den})$'' in this work; i.e.
\begin{equation}
 \tan\gamma({\rm den})
 =-{\displaystyle\frac{\sqrt{2}\langle \hat{Q}_{22} \rangle}
 {\langle \hat{Q}_{20} \rangle}}.
\label{eq:gammaden}
\end{equation}
Since $\gamma(\mbox{den})$ depends on the calculated single-particle
wave functions of the average nuclear potential, and on the configuration
of each nucleus, it is more convenient
to introduce an another parameter ``$\gamma(\mbox{geo})$'', which is
more directly related to the geometric shape of nucleus mathematically
defined by the two dimensional surface $\Sigma$:
\begin{equation}
 \tan\gamma(\mbox{geo})
 =-{\displaystyle\frac{\sqrt{2}\langle \hat{Q}_{22} \rangle_{\rm uni}}
 {\langle \hat{Q}_{20} \rangle_{\rm uni}}},
\label{eq:gammageo}
\end{equation}
where $\langle\quad\rangle_{\rm uni}$ means that the expectation
value is taken with respect to the sharp-cut uniform density distribution,
\begin{equation}
 \rho_{\rm uni}(\mbox{\boldmath$r$})\equiv\left\{
 \begin{array}{ll} \rho_0 \vspace{1mm}&
 \mbox{for \mbox{\boldmath$r$} inside the surface $\Sigma$},\\
 0 & \mbox{otherwise}.\end{array}\right.
\label{eq:rhouni}
\end{equation}

\begin{figure}[htbp]
\includegraphics[width=8cm,keepaspectratio]{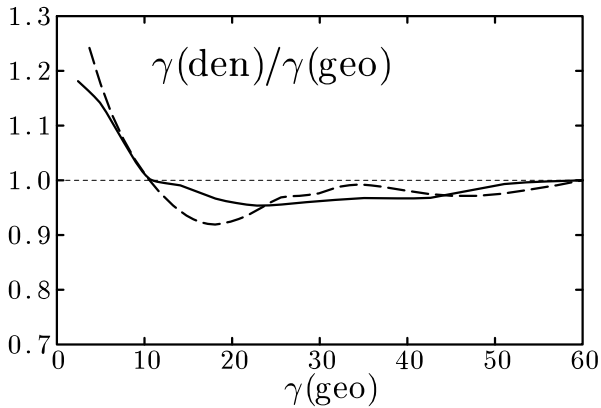}
\caption{
 The ratios $\gamma(\mbox{den})/\gamma(\mbox{geo})$ are plotted
 as functions of $\gamma(\mbox{geo})$ for the cases of
 the Nilsson (solid line) and Woods-Saxon potentials (dashed line).
 The calculations have been performed for the TSD band in $^{163}$Lu,
 and the parameters used for the Nilsson potential are
 $\epsilon_2=0.43$, $\epsilon_4=0.0$, and ${\mit\Delta}_{\rm n,p}=0.3$ MeV,
 and those for the Woods-Saxon potential are
 $\beta_2=0.42$, $\beta_4=0.034$, and ${\mit\Delta}_{\rm n,p}=0.3$ MeV,
 which has almost the same shape as the case of the Nilsson potential
 at $\gamma(\mbox{geo})\approx 10^\circ$
 in the minimum of the Nilsson-Strutinsky calculation.
}
\label{fig:gamcons}
\end{figure}

If the density is calculated within the mean-field approximation
by using the average single-particle potential
which has the same shape as the uniform density~(\ref{eq:rhouni}),
then these two triaxiality parameters,
$\gamma(\mbox{den})$ and $\gamma(\mbox{geo})$, agree very well.
In Fig.~\ref{fig:gamcons}
the ratio $\gamma(\mbox{den})/\gamma(\mbox{geo})$
is shown as a function of $\gamma(\mbox{geo})$, and we can see
that they coincide typically within 10\% except in the small $\gamma$
region, where both $\gamma(\mbox{den})$ and $\gamma(\mbox{geo})$ become
zero and the ratio is numerically unstable.
This agreement corresponds to the so-called shape consistency
between the density and the potential~\cite{NPP77},
which has been tested both for the Nilsson and Woods-Saxon potentials
for the axially symmetric deformations~\cite{DNO84,BDNO89}.
Therefore, we can practically use $\gamma(\mbox{geo})$
in place of $\gamma(\mbox{den})$.

The definition of triaxiality parameter,
$\gamma(\mbox{den})$ or $\gamma(\mbox{geo})$, is basic or
fundamental in the sense that it is directly related to
the $B(E2)$ values of the triaxial rotor.
In the microscopic calculations by means of the Hartree-Fock
or Hartree-Fock-Bogoliubov method, $\gamma=\gamma(\mbox{den})$
is the only definition of the triaxial deformation, and no confusion exists.
It is, however, quite often that one starts from some average nuclear potential,
which is triaxially deformed, and calculates
the potential energy surface or the total routhian surface
to determine the selfconsistent deformation
as in the case of the Strutinsky method.
In such cases, a different type of definitions of the triaxiality
parameter $\gamma$ has been used,
which is chosen conveniently to specify
the triaxial deformation of the potential.
We call this third type of definition ``$\gamma(\mbox{pot})$'':
We consider two well-known definitions in the followings,
depending on the employed potential.

\subsection{\mbox{\boldmath$\gamma$}(pot) defined in the Nilsson potential}
\label{sect:gamNils}

As a definite example of $\gamma(\mbox{pot})$,
we take the Nilsson potential, or the modified oscillator potential,
as a mean-field potential,
i.e. $\gamma(\mbox{pot:Nils})$.  The deformation parameters
in the Nilsson potential~\cite{NTS69,BR85,NR95} considered
in the present work are $(\epsilon_2,\gamma,\epsilon_4)$,
which define the deformation of
the velocity independent part of potential through
the single-stretched coordinate,
$\mbox{\boldmath$r$}'\equiv
(\sqrt{\omega_x/\omega_0}\,x,\sqrt{\omega_y/\omega_0}\,y,
\sqrt{\omega_z/\omega_0}\,z)$, as
\begin{equation}
\begin{array}{l}
{\displaystyle
V(\mbox{\boldmath$r$})=
\frac{1}{2}M\omega_0\omega_{\rm v}(\epsilon_2,\gamma,\epsilon_4) r'^2 \times
 }\vspace{1mm} \\
{\displaystyle
\quad
\Bigl(1 -\sum_{K=0,\pm 2}c_{2K}Y_{2K}(\Omega')
 -\sum_{K=0,\pm 2, \pm 4}c_{4K}Y_{4K}(\Omega')
 \Bigr),
 }
\end{array}
\label{eq:Nilsdef}
\end{equation}
where $\omega_0$ is the frequency of the spherical potential,
$\omega_{\rm v}(\epsilon_2,\gamma,\epsilon_4)$ is determined
by the volume conserving condition, $\Omega'$ is the solid-angle
of coordinate $\mbox{\boldmath$r$}'$,
and the coefficients $c$'s are given by
\begin{equation}
\left\{ \begin{array}{l}
 c_{20}=\sqrt{\frac{16\pi}{45}}\,\epsilon_2\cos\gamma,\\
 c_{22}=c_{2-2}=-\sqrt{\frac{8\pi}{45}}\,\epsilon_2\sin\gamma,\\
 c_{40}=\frac{\sqrt{4\pi}}{9}\,\epsilon_4
  (5\cos^2\gamma+1),\\
 c_{42}=c_{4-2}=-\frac{\sqrt{120\pi}}{9}\,\epsilon_4
  \cos\gamma\sin\gamma,\\
 c_{44}=c_{4-4}=\frac{\sqrt{70\pi}}{9}\,\epsilon_4
  \sin^2\gamma.
\end{array} \right.
\label{eq:Nilseps}
\end{equation}
Note that the three frequencies, $\omega_x$, $\omega_y$, and $\omega_z$,
are given by Eq.~(\ref{eq:OMLund}) below.
The nuclear shape $\Sigma$ in this case is defined
as an equi-potential surface of the potential,
$V(\mbox{\boldmath$r$})=$ const., and is uniquely determined
once the parameters  $(\epsilon_2,\gamma,\epsilon_4)$ with
$\gamma=\gamma(\mbox{pot:Nils})$ are given.
It is straightforward but rather complicated because of
the use of the single-stretched coordinate $\mbox{\boldmath$r$}'$ in practice.
Using this potential either the triaxiality parameter
$\gamma(\mbox{den})$ or $\gamma(\mbox{geo})$
defined in the previous subsection
can be calculated as functions of these potential parameters
$(\epsilon_2,\gamma=\gamma(\mbox{pot:Nils}),\epsilon_4)$.

\begin{figure}[htbp]
\includegraphics[width=6cm,keepaspectratio]{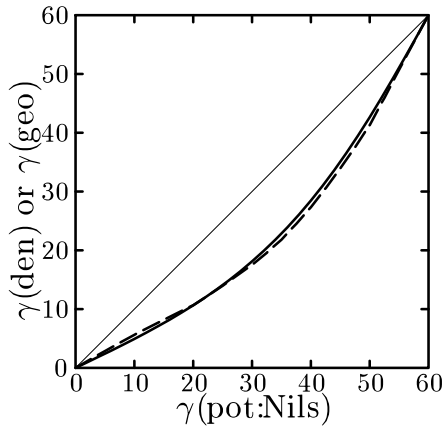}
\caption{
 The triaxiality parameters $\gamma(\mbox{den})$ (dashed line)
 and $\gamma(\mbox{geo})$ (solid line)
 are shown as functions of the $\gamma$ parameter in the Nilsson
 potential, $\gamma(\mbox{pot:Nils})$
 (the thin diagonal line is just for a guide to the eyes).
 The calculations have been performed for the TSD band in $^{163}$Lu,
 and the parameters used are
 $\epsilon_2=0.43$, $\epsilon_4=0.0$, and ${\mit\Delta}_{\rm n,p}=0.3$ MeV.
}
\label{fig:gamNl}
\end{figure}

\begin{figure}[htbp]
\includegraphics[width=6cm,keepaspectratio]{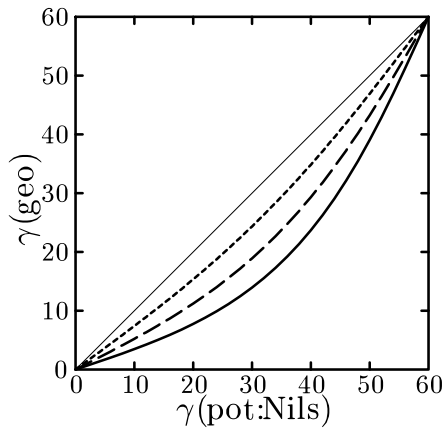}
\caption{
 The triaxiality parameters $\gamma(\mbox{geo})$
 with fixed $\epsilon_2$ and $\epsilon_4$ deformation parameters
 are shown as functions of $\gamma(\mbox{pot:Nils})$
 (the thin diagonal line is just for a guide to the eyes).
 The dotted line is for the case with $\epsilon_2=0.2$,
 the dashed with $\epsilon_2=0.4$, and
 the solid with $\epsilon_2=0.6$, respectively.
 The $\epsilon_4$ parameter is set 0 for all the cases.
}
\label{fig:gamnla}
\end{figure}

In Fig.~\ref{fig:gamNl}, are depicted the relation
between $\gamma(\mbox{den})$
and $\gamma(\mbox{pot:Nils})$ and that between $\gamma(\mbox{geo})$
and $\gamma(\mbox{pot:Nils})$ at a given $(\epsilon_2,\epsilon_4)$,
which is suitable for the TSD band in $^{163}$Lu.
As is already shown in Fig.~\ref{fig:gamcons},
$\gamma(\mbox{den})$ and $\gamma(\mbox{geo})$
are very similar, but $\gamma(\mbox{pot:Nils})$ is quite different:
$\gamma(\mbox{pot:Nils})=20^\circ$ corresponds to
$\gamma(\mbox{den})\approx\gamma(\mbox{geo})\approx 11^\circ$,
so that the difference can be as much as about a factor two.
In order to see how the difference between $\gamma(\mbox{geo})$
and $\gamma(\mbox{pot:Nils})$ changes for different $\epsilon_2$ values,
three cases with $\epsilon_2=0.2$, 0.4, 0.6 are shown in Fig.~\ref{fig:gamnla}.
It is clear that the difference becomes larger
for larger $\epsilon_2$ deformations: $\gamma(\mbox{geo})$ is only
about $10^\circ$ even though $\gamma(\mbox{pot:Nils})$ is put $30^\circ$
in the case of the superdeformed band, $\epsilon_2\approx 0.6$.

In the case of $\epsilon_4=0$, the Nilsson potential reduces to
the anisotropic harmonic oscillator potential
except for the $\mbox{\boldmath$l$}^2$ and
$\mbox{\boldmath$l$}\cdot\mbox{\boldmath$s$}$ terms,
which are irrelevant for the definition of the nuclear shape.
It is instructive to consider such a case in order to understand
the difference shown in Fig.~\ref{fig:gamnla}.
Then the shape is a volume-conserving ellipsoid defined by simple equations,
\begin{equation}
 \sum_{i=1}^{3} \omega_i^2 x_i^2 = \mbox{const.},\quad\mbox{with}\quad
 \prod_{i=1}^{3} \omega_i=\omega_0^3.
\label{eq:HOpot}
\end{equation}
The frequencies $\omega_i$ ($i=1,2,3$) for the $x$, $y$, $z$-directions,
which are inversely proportional to the lengths of the ellipsoid
along these axes, are given by
\begin{equation}
 \omega_i = \omega_{\rm v}\left(1-\frac{2}{3}\epsilon_2
 \cos\Bigl(\gamma+\frac{2\pi}{3}i\Bigr)\right),
 \quad \gamma=\gamma(\mbox{pot:Nils}).
\label{eq:OMLund}
\end{equation}
Therefore $\gamma(\mbox{geo})$ and $\gamma(\mbox{pot:Nils})$
are related through $\omega_i$ ($i=1,2,3$);
\begin{eqnarray}
 &&\tan\gamma(\mbox{geo})=
 {\displaystyle
  \frac{\sqrt{3}\,(\omega_y^{-2}-\omega_x^{-2})}
        {2\omega_z^{-2}-\omega_y^{-2}-\omega_x^{-2}}
\label{eq:HOgeo}
  },\\
 &&\tan\gamma(\mbox{pot:Nils})=
 {\displaystyle
  \frac{\sqrt{3}\,(\omega_y-\omega_x)}
        {2\omega_z-\omega_y-\omega_x}
\label{eq:HOpotNils}
  },
\end{eqnarray}
which relate these two $\gamma$'s for a given value of $\epsilon_2$.
In the limit of small deformation parameters, $\epsilon_2$,$|\gamma|\ll 1$,
it is easy to confirm
\begin{equation}
 \gamma(\mbox{geo}) \approx \Bigl(1-\frac{3}{2}\epsilon_2\Bigr)\,
 \gamma(\mbox{pot:Nils}).
\label{eq:geoNils}
\end{equation}
Namely, the slope of curves at the origin in Fig.~\ref{fig:gamnla}
changes with $\epsilon_2$ with a rather large factor $\frac{3}{2}$,
and this clearly explains that $\gamma(\mbox{geo})$ is smaller than
$\gamma(\mbox{pot:Nils})$ more than a factor two when $\epsilon_2$ is as large
as 0.4 like in the case of the TSD band.

\begin{figure}[htbp]
\includegraphics[width=75mm,keepaspectratio]{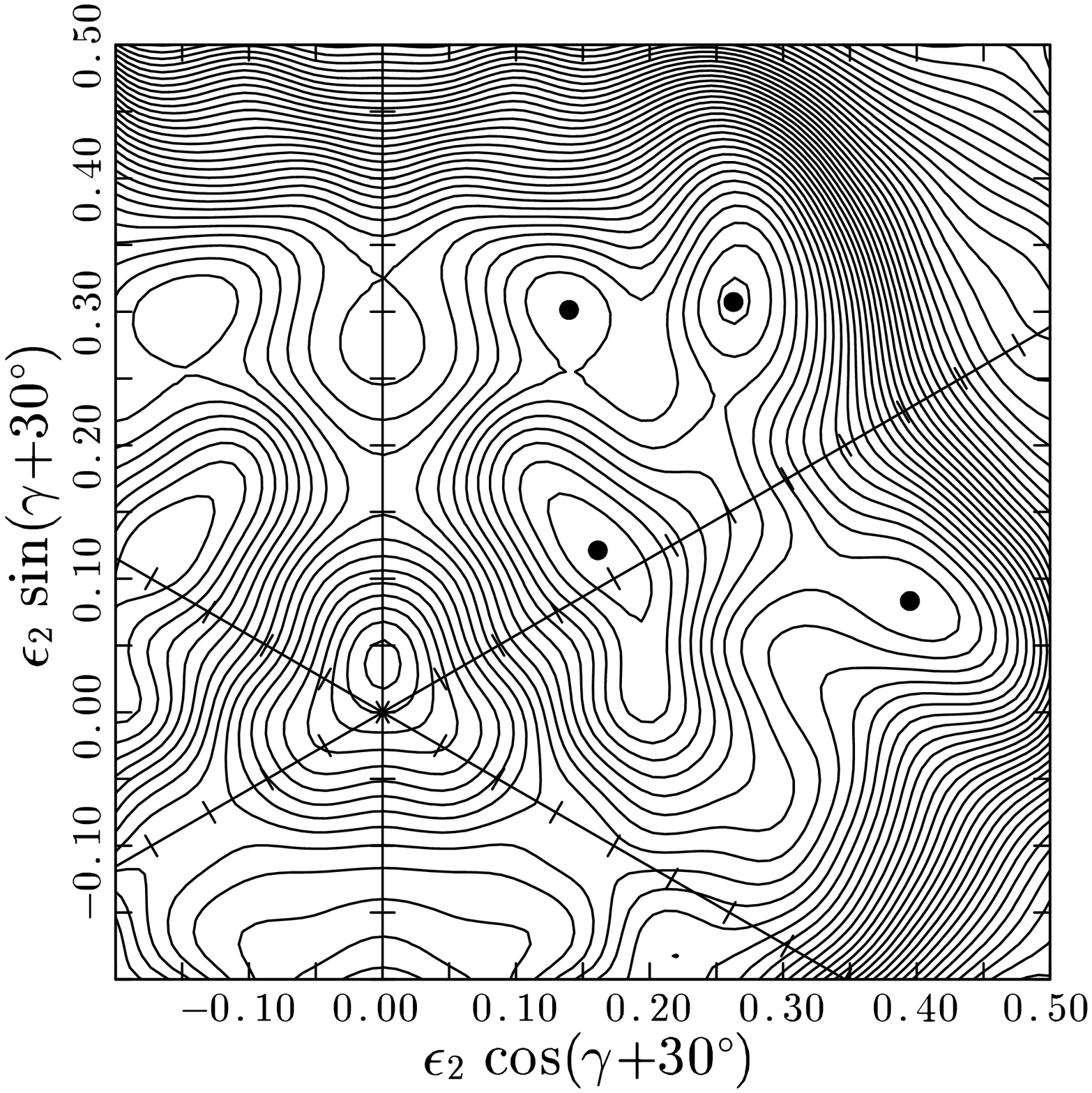}\\
\includegraphics[width=75mm,keepaspectratio]{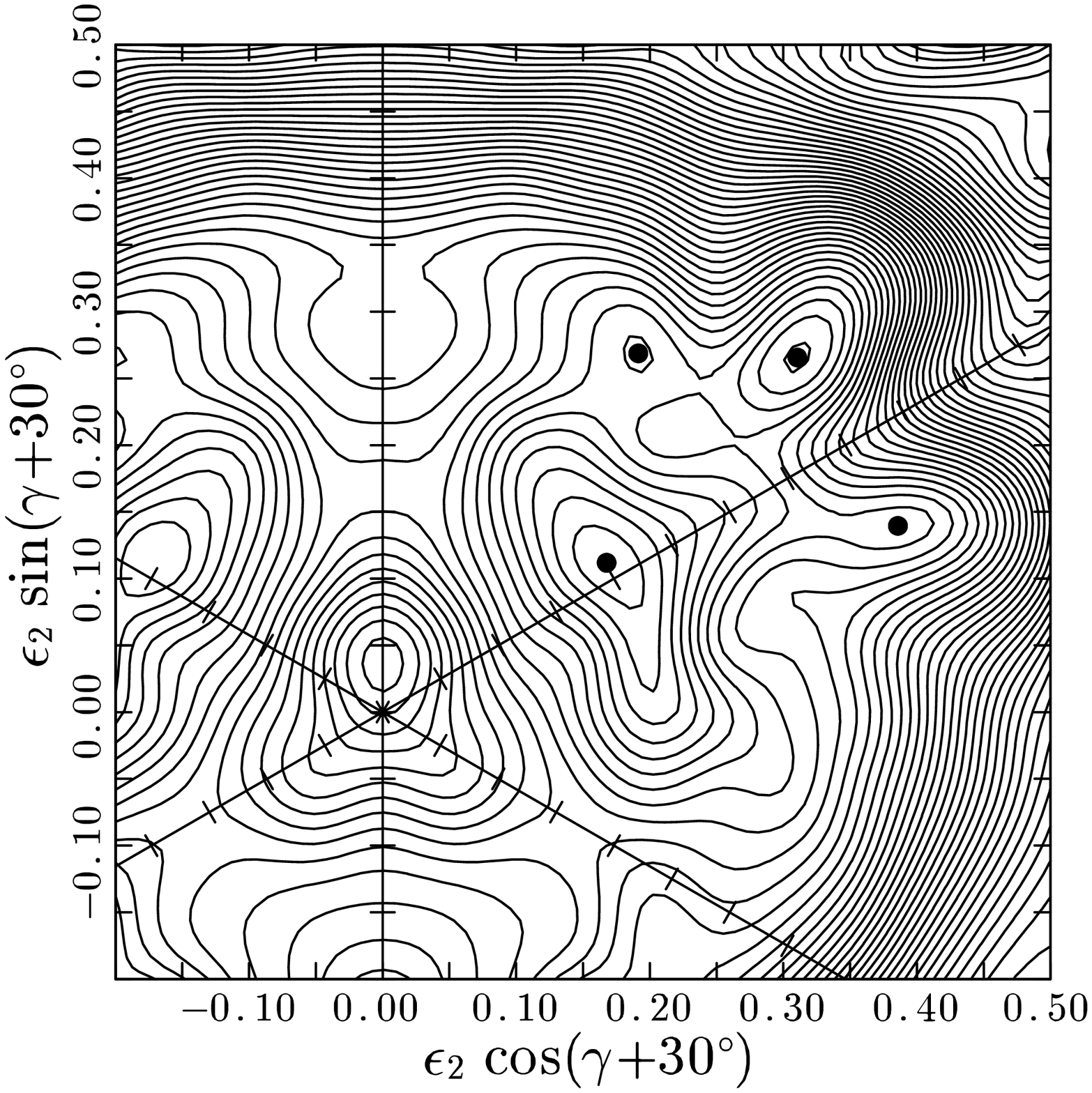}
\caption{
 Potential energy surface obtained by
 the cranked Nilsson Strutinsky calculation for
 the $(\pi,\alpha)=(+,+1/2)$ configuration in $^{163}$Lu at $I=41/2^+$.
 The energy between contours is 250 keV.
 The triaxiality parameter $\gamma=\gamma(\mbox{pot:Nils})$
 is used as usual in the upper panel,
 while the triaxiality parameter $\gamma=\gamma(\mbox{geo})$
 in the lower panel.
}
\label{fig:potsurf}
\end{figure}

In order to see how these different definitions of two triaxiality
parameters, $\gamma(\mbox{geo})$ and $\gamma(\mbox{pot:Nils})$,
change the appearance of potential energy surface,
we show an example in Fig.~\ref{fig:potsurf}.
Here the $\epsilon_4$ parameter is chosen to minimize the potential energy
at each $(\epsilon_2,\gamma)$ mesh points.
The parameter $\gamma(\mbox{geo})$ depends not only
on $(\epsilon_2,\gamma(\mbox{pot:Nils}))$ but on $\epsilon_4$,
and it is impossible to calculate the $(\epsilon_2,\gamma(\mbox{geo}))$
mesh points before the minimization with respect to $\epsilon_4$.
Therefore, we made an approximation to set $\epsilon_4=0$
when we prepare the $(\epsilon_2,\gamma(\mbox{geo}))$
mesh points from the $(\epsilon_2,\gamma(\mbox{pot:Nils}))$ mesh points.
As is clear from the figure, the surface is squeezed to
the $\gamma=0$ axis at larger deformation, and apparently
the TSD minimum moves to smaller triaxial values.
In Fig.~\ref{fig:potsurf},
only the $\gamma$ parameter is replaced
from $\gamma(\mbox{pot:Nils})$ to $\gamma(\mbox{geo})$.
However, it may be better to replace $\epsilon_2$ to
the other parameter corresponding to the magnitude $Q$
in Eq.(\ref{eq:Qgamma}) in order to make the meaning of
the quadrupole deformation clearer.
Constraint Hartree-Fock(-Bogoliubov) type calculations are necessary
for such a purpose.  Then the method becomes much more involved
and the simplicity of the Strutinsky type calculation may be lost.

It may be worthwhile mentioning, here, the specific model composed
of the spherical Nilsson potential and the $QQ$ force as an effective
interaction.  The velocity-independent part of the Hartree potential
in this model is given by
\begin{equation}
\begin{array}{l}
{\displaystyle
V(\mbox{\boldmath$r$})=
\frac{1}{2}M\omega_0^2\,\mbox{\boldmath$r$}^2
 -\alpha_{20}{\hat Q}_{20}
 -\alpha_{22}({\hat Q}_{22}+{\hat Q}_{2-2}),
 }
\end{array}
\label{eq:NilsQQ}
\end{equation}
where the Hartree condition requires
$\alpha_{2K}=\chi\,\langle{\hat Q}_{2K}\rangle$ ($K=0,2$)
with $\chi$ being the $QQ$ force strength,
and the potential reduces also to the anisotropic harmonic oscillator
(but the volume conservation condition is not necessarily satisfied
in this Hartree procedure).
If the model space is not restricted, $\alpha_{2K}$ is proportional
to the two intrinsic quadrupole moments in Eq.~(\ref{eq:Qintr}),
and then, the triaxiality parameter of the density type~(\ref{eq:gammaden})
in this case is related to $-\sqrt{2}\,\alpha_{22}/\alpha_{20}$,
which can be expressed in terms of $\omega_i$ ($i=1,2,3$) as
\begin{equation}
 \tan\gamma(\mbox{den:QQ})=
 {\displaystyle
  \frac{\sqrt{3}\,(\omega_y^{2}-\omega_x^{2})}
      {2\omega_z^{2}-\omega_y^{2}-\omega_x^{2}}
 }.
\label{eq:HOQQ}
\end{equation}
It is usual to parametrize the potential by the deformation
parameter $(\delta,\gamma)$ with $\gamma=\gamma(\mbox{den:QQ}))$,
in terms of which the frequencies are written in the form,
\begin{equation}
 \omega_i^2 = \omega_0^2\left(1-\frac{4}{3}\delta
 \cos\Bigl(\gamma+\frac{2\pi}{3}i\Bigr)\right),
 \quad \gamma=\gamma(\mbox{den:QQ}).
\label{eq:OMQQ}
\end{equation}
Thus, the triaxiality parameters $\gamma(\mbox{geo})$ and
$\gamma(\mbox{den:QQ})$ are quite different in this case:
In the limit of small deformation parameters, $\delta$,$|\gamma|\ll 1$,
they are related like
\begin{equation}
 \gamma(\mbox{geo}) \approx (1-2\delta)\,
 \gamma(\mbox{den:QQ}).
\label{eq:geoQQ}
\end{equation}
Namely, the difference is even larger than that between
$\gamma(\mbox{geo})$ and $\gamma(\mbox{pot:Nils})$.
This means that the shape consistency between
the density and the potential is strongly violated in the $QQ$ force model,
if the full model space is used in the Hartree procedure.
Considering this on top of the fact that the volume conservation
condition is not guaranteed, the model is not realistic at all.

Actually, the $QQ$ force model is supposed to be a model
in a restricted model space, and the contributions of the ``core''
should be added to $\langle{\hat Q}_{2K}\rangle$ calculated
within the model space in order to obtain the intrinsic moments
of the whole system.
Then the problem of breaking the shape consistency
may not be the real problem.
In fact, the core contributions of the quadrupole operators,
i.e. the expectation values for the closed shell configurations
in the anisotropic harmonic oscillator potential,
have different dependence on the deformation;
it can be shown~\cite{SM84IVD} that
they lead to the triaxiality parameter $\gamma=\gamma(\mbox{den:core})$,
which satisfies
\begin{equation}
 \tan\gamma(\mbox{den:core})=
 {\displaystyle
  \frac{\sqrt{3}\,(\omega_y^{-1}-\omega_x^{-1})}
      {2\omega_z^{-1}-\omega_y^{-1}-\omega_x^{-1}}
 }.
\label{eq:HOcore}
\end{equation}
In Ref.~\cite{SM84IVD}
(see Appendix B of this reference for details,
but note the different notations used there),
these differences between various types
of the deformation parameters defined in the harmonic
oscillator potential were already discussed,
where not only the parameter $\gamma$ but also the other one of a pair
of the parameters,
$\beta=
\sqrt{\frac{4\pi}{5}}\,Q/ \langle \sum_{k=1}^{A}\mbox{\boldmath$r$}^2\rangle$,
were considered.
It was already pointed out that the triaxiality parameters
for a given shape in various types of definition take quite different values.

\subsection{\mbox{\boldmath$\gamma$}(pot) defined in the Woods-Saxon potential}
\label{sect:gamWS}

As an another example of $\gamma(\mbox{pot})$, the parametrization of
deformation in the Woods-Saxon potential is considered,
i.e. $\gamma(\mbox{pot:WS})$.  Actually,
it is not restricted to the Woods-Saxon potential, but is more general
as one can see in the following.  The deformed Woods-Saxon potential
considered in this work is parametrized by the deformation parameters,
$(\beta_2,\gamma,\beta_4)$, and defined~\cite{DMS79,RS81,NR81} by
\begin{equation}
{\displaystyle
 V(\mbox{\boldmath$r$})=
 \frac{V_0}{1+\exp(\mbox{dist}_\Sigma(\mbox{\boldmath$r$})/a)}
 },
\label{eq:WSdef}
\end{equation}
where $\mbox{dist}_\Sigma(\mbox{\boldmath$r$})$ is the distance
between a given point $\mbox{\boldmath$r$}$
and the nuclear surface $\Sigma$,
with a minus sign if $\mbox{\boldmath$r$}$ is inside $\Sigma$,
which is defined by the usual radius to solid-angle relation,
$r=R(\Omega)$;
\begin{equation}
\begin{array}{l}
{\displaystyle
 R(\Omega)=R_{\rm v}(\beta_2,\gamma,\beta_4)\times
 }\vspace{1mm} \\
{\displaystyle
\quad
\Bigl(1 +\sum_{K=0,\pm 2}a_{2K}Y_{2K}(\Omega)
 +\sum_{K=0,\pm 2, \pm 4}a_{4K}Y_{4K}(\Omega)
 \Bigr),
 }
\end{array}
\label{eq:defRS}
\end{equation}
where $R_{\rm v}(\beta_2,\gamma,\beta_4)$ is determined
by the volume conserving condition,
and the coefficients $a$'s are given by
\begin{equation}
\left\{ \begin{array}{l}
 a_{20}=\beta_2\cos\gamma,\\
 a_{22}=a_{2-2}=-\frac{1}{\sqrt{2}}\,\beta_2\sin\gamma,\\
 a_{40}=\frac{1}{6}\,\beta_4
  (5\cos^2\gamma+1),\\
 a_{42}=a_{4-2}=-\sqrt{\frac{5}{6}}\,\beta_4
  \cos\gamma\sin\gamma,\\
 a_{44}=a_{4-4}=\sqrt{\frac{35}{72}}\,\beta_4
  \sin^2\gamma.
\end{array} \right.
\label{eq:WSbeta}
\end{equation}
Apparently the surface $\Sigma$ is given as an equi-potential
surface at the half depth, $V(\mbox{\boldmath$r$})=\frac{1}{2}V_0$,
and it is directly related to $(\beta_2,\gamma,\beta_4)$ with
$\gamma=\gamma(\mbox{pot:WS})$.

\begin{figure}[htbp]
\includegraphics[width=6cm,keepaspectratio]{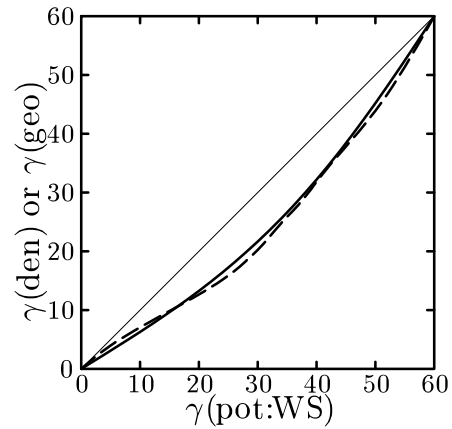}
\caption{
 The triaxiality parameters $\gamma(\mbox{den})$ (dashed line)
 and $\gamma(\mbox{geo})$ (solid line)
 are shown as functions of the $\gamma$ parameter in the Woods-Saxon
 potential, $\gamma(\mbox{pot:WS})$
 (the thin diagonal line is just for a guide to the eyes).
 The calculations have been performed for the TSD band in $^{163}$Lu,
 and the parameters used are
 $\beta_2=0.42$, $\beta_4=0.034$, and ${\mit\Delta}_{\rm n,p}=0.3$ MeV;
 these deformation parameters gives almost the same shape
 as that used in the Nilsson potential in Fig.~\ref{fig:gamNl}
 at $\gamma(\mbox{geo})\approx 10^\circ$.
}
\label{fig:gamWS}
\end{figure}

\begin{figure}[htbp]
\includegraphics[width=6cm,keepaspectratio]{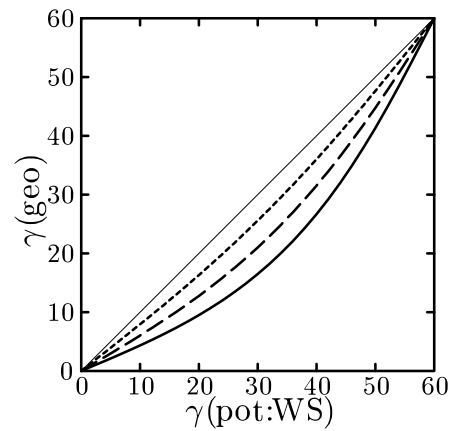}
\caption{
 The triaxiality parameters $\gamma(\mbox{geo})$
 with fixed $\beta_2$ and $\beta_4$ deformation parameters
 are shown as functions of $\gamma(\mbox{pot:WS})$
 (the thin diagonal line is just for a guide to the eyes).
 The dotted line is for the case with $\beta_2=0.217$ and $\beta_4=0.017$,
 the dashed with $\beta_2=0.445$ and $\beta_4=0.075$, and
 the solid with $\beta_2=0.685$ and $\beta_4=0.190$, respectively.
 These sets of parameters almost correspond to the cases of
 $\epsilon_2=0.2$, $\epsilon_2=0.4$, and $\epsilon_2=0.6$
 with $\epsilon_4=0$ used
 for the Nilsson potential in Fig.~\ref{fig:gamnla}
 at $\gamma(\mbox{geo})=0$.
}
\label{fig:gamwsa}
\end{figure}

As in the case of the Nilsson potential,
the triaxiality parameter
$\gamma(\mbox{den})$ and $\gamma(\mbox{geo})$ are shown as functions
of $\gamma(\mbox{pot:WS})$ in Fig.~\ref{fig:gamWS},
which are suitable to the TSD band in $^{163}$Lu,
just in the same way as in the case
of the Nilsson potential in Fig.~\ref{fig:gamNl}.
Again, $\gamma(\mbox{den})$ and $\gamma(\mbox{geo})$ are very similar,
but they are quite different from $\gamma(\mbox{pot:WS})$:
$\gamma(\mbox{den})\approx\gamma(\mbox{geo})\approx 13^\circ$ when
$\gamma(\mbox{pot:WS})=20^\circ$.
In Fig.~\ref{fig:gamwsa}, the relation between $\gamma(\mbox{geo})$
and $\gamma(\mbox{pot:WS})$ at three different cases of
$(\beta_2,\beta_4)$ deformations, corresponding to Fig.~\ref{fig:gamnla},
are also depicted.  Although the differences between $\gamma(\mbox{geo})$
and $\gamma(\mbox{pot:WS})$ are not so dramatic as
those between $\gamma(\mbox{geo})$ and $\gamma(\mbox{pot:Nils})$,
they are still considerably large.
In the case of the parametrization
of the nuclear surface in Eq.~(\ref{eq:defRS}),
$\langle{\hat Q}_{2K}\rangle_{\rm uni}$ can be easily calculated;
\begin{equation}
\langle{\hat Q}_{2K}\rangle_{\rm uni}
 =\frac{1}{5}\int R(\Omega)^5 Y_{2K}(\Omega)d\Omega.
\label{eq:QuniWS}
\end{equation}
Then, it is straightforward to see, in the small deformation limit,
$\beta_2$,$|\gamma|\ll 1$, with $\beta_4=0$, that
\begin{equation}
 \gamma(\mbox{geo}) \approx
 \Bigl(1-\sqrt{\frac{180}{49\pi}}\,\beta_2\Bigr)\,
 \gamma(\mbox{pot:WS}).
\label{eq:geoWS}
\end{equation}
Taking into account the relation,
$\beta_2 \approx \sqrt{\frac{16\pi}{45}}\,\epsilon_2$
in the small deformation limit,
the proportionality constant in front of $\epsilon_2$
corresponds, in this case, to
$\sqrt{\frac{180}{49\pi}}\times\sqrt{\frac{16\pi}{45}}
=\frac{8}{7}\approx 1.14$, which is smaller than $\frac{3}{2}=1.5$
in Eq.~(\ref{eq:geoNils}) for the Nilsson potential,
but is still appreciably large.
This explains qualitatively the increase of the difference between
$\gamma(\mbox{geo})$ and $\gamma(\mbox{pot:WS})$
for larger deformations as is shown in Fig.~\ref{fig:gamwsa}.

\section{\mbox{\boldmath $B(E2)$} ratio of the wobbling band}
\label{sect:be2ratio}

As it is discussed in the previous section,
the two intrinsic quadrupole moments should be determined
in order to deduce the triaxial deformation.
In the case of the wobbling excitations, it is enough to measure
the two $B(E2)$'s, $B(E2)_{\rm in}$ and $B(E2)_{\rm out}$;
the ${\mit\Delta}I=-2$ intraband $E2$ transitions within the wobbling band
and ${\mit\Delta}I=\pm 1$ interband $E2$ transitions from
the one-phonon wobbling band to the yrast TSD band, respectively.
According to the rotor model~\cite{BM75},
the magnitude of the moment $Q$ in Eq.(\ref{eq:Qgamma}) is
factored out in the two $B(E2)$'s and their ratio,
$B(E2)_{\rm out}/B(E2)_{\rm in}$, is directly
related to the triaxiality parameter $\gamma(\mbox{den})$.
This $B(E2)$ ratio is straightforward to measure
from the experimental point of view;
it is given directly by the $\gamma$-ray branching ratio
if the information of the mixing ratio is provided.
In contrast, the life time measurement is necessary
to obtain $B(E2)$ values themselves, which is not an easy task generally.
Although the life time measurements have been done
in some TSD bands~\cite{wobLuQ,wob163LuQ} recently,
so that we can study both $B(E2)_{\rm out}$ and $B(E2)_{\rm in}$
separately, we concentrate upon the $B(E2)$ ratio
in the present work.

\begin{figure}[htbp]
\includegraphics[width=7cm,keepaspectratio]{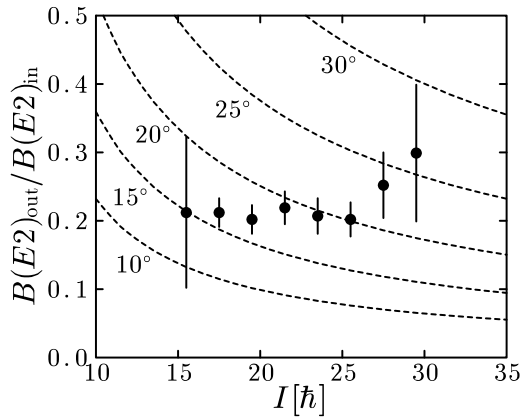}
\caption{
 The $B(E2)$ ratio,
 $B(E2\mbox{:}\,I\rightarrow I-1)_{\rm out}/
 B(E2\mbox{:}\,I\rightarrow I-2)_{\rm in}$,
 of the wobbling band in $^{163}$Lu.
 The experimental data~\cite{wob163LuQ} is compared with
 the calculations by the particle-rotor model~\cite{HH03}.
 Five dotted curves are the results with the triaxiality parameter
 $\gamma(\mbox{den})=10^\circ$, $15^\circ$, $20^\circ$,
 $25^\circ$, and $30^\circ$ from the lowest to the highest ones,
 respectively.
}
\label{fig:rotBE2r}
\end{figure}

In Fig.~\ref{fig:rotBE2r}, the experimental $B(E2)$ ratio
of the one-phonon wobbling band in $^{163}$Lu~\cite{wob163LuQ}
is compared with the results of the particle-rotor model
calculation in Ref.~\cite{HH03}.
Important parameters of the model are three moments of inertia,
${\cal J}_x$, ${\cal J}_y$,  ${\cal J}_z$,
and the triaxiality $\gamma=\gamma(\mbox{den})$;
an overall factor of the formers is irrelevant to the $B(E2)$ ratio
and they are fixed to be ${\cal J}_x:{\cal J}_y:{\cal J}_z=145:135:50$
taken from~\cite{HH03},
while we take five values $\gamma(\mbox{den})=10^\circ$,
$15^\circ$, $20^\circ$, $25^\circ$, and $30^\circ$ in order to
show the dependence of the $B(E2)$ ratio on the $\gamma$ values.
Other parameters, the chemical potential of the odd $i_{13/2}$ proton,
$\lambda/\kappa=-1.532$, and the pairing gap, $\Delta/\kappa=0.3$,
with $\kappa=3$ MeV are also taken from~\cite{HH03}.
The calculated ratios are monotonically decreasing functions of spin
if all the model parameters are held fixed.
This decrease is characteristic in the rotor model,
see Eq.~(\ref{eq:BE2rot}) below.
Although the spin-dependence is somewhat different,
the average value of the $B(E2)$ ratio can be reproduced
if we take the value $\gamma\approx 20^\circ$.
Therefore the triaxial deformation of $^{163}$Lu is deduced
to be $\gamma\approx 20^\circ$, which is one of the main conclusions
of Ref.~\cite{HH03}.  This conclusion remains valid even if the parameters
of the moments of inertia are changed in a reasonable range.
While the existence of the odd proton brings about important corrections
to the energy spectra,
its effect on the $B(E2)$ is very small~\cite{HH03,TST06}.
Thus, the wobbling phonon treatment of the simple rotor model
in Ref.~\cite{BM75} gives a good approximation to the $B(E2)$ ratio
at high-spin states, leading to the following expression;
\begin{equation}
\begin{array}{l}
{\displaystyle
 \frac{B(E2\mbox{:}\,I\rightarrow I\pm 1)_{\rm out}}
      {B(E2\mbox{:}\,I\rightarrow I-2)_{\rm in}}} \vspace{1mm}\\
 \quad\approx
{\displaystyle
 \frac{2}{I}\left(
 \frac{w_z\sin(\gamma+60^\circ)\mp w_y\sin{\gamma}}
 {\sqrt{w_y w_z}\cos(\gamma+30^\circ)}\right)^2,
 \quad \gamma=\gamma(\mbox{den})},
\end{array}
\label{eq:BE2rot}
\end{equation}
where the quantities $w_y,w_z$ are related to
the three moments of inertia through
\begin{equation}
\left\{\begin{array}{l}
w_y\equiv ({\cal J}_x/{\cal J}_z - 1)^{1/2},\\
w_z\equiv ({\cal J}_x/{\cal J}_y - 1)^{1/2}.
\end{array}\right.
\label{eq:BE2rotw}
\end{equation}
Note that the $I\rightarrow I+1$ transitions are quenched for the positive
$\gamma$ shape, and in fact only the $I\rightarrow I-1$ transitions
are observed in experiment in the Lu nuclei.
It is easy to check that the results of calculations
in Fig.~\ref{fig:rotBE2r} can be understood nicely
by this simple expression;
the decrease of the $B(E2)$ ratio as a function of spin
is due to the $1/I$ dependence in Eq.(\ref{eq:BE2rot}),
and the ratio increases quickly as a function of
the triaxiality $\gamma=\gamma(\mbox{den})$.

It may be interesting to note that
the measured $B(E2)$ ratio is almost constant as a function of spin
or even increases at highest spins, which is quite different
from that of the rotor model calculation.
It indicates that the parameters of the model are
changing as the spin $I$ increases.
Since the ratio is sensitive to the triaxiality parameter
$\gamma=\gamma(\mbox{den})$, it is natural to consider
that $\gamma(\mbox{den})$ is spin dependent and
increases with spins~\cite{HH03} among others.
A preliminary investigation for such a possibility
has been reported in Ref.~\cite{SS06},
where is employed a microscopic framework, the cranked Woods-Saxon mean-field
and the random phase approximation (RPA).

Next let us turn to the discussion on our microscopic calculations
in Refs.~\cite{MSMwob1,MSMwob2,MSMwob3}, which are based on
the cranked Nilsson mean-field and the random phase approximation (RPA).
In these calculations the triaxiality parameter $\gamma=20^\circ$
was employed but the resultant $B(E2)$ ratios were too small
by a factor two to three.
We were wondering about possible reasons;
does the results of RPA calculation deviate from those of
the rotor model so much?  However, it has been shown in Ref.~\cite{SMMprec}
that the RPA calculation reproduces the result of the rotor model
rather well in the case of the precession bands, which are nothing but
the rotational bands built upon the high-$K$ isomers and
can be interpreted as a similar motion to the wobbling excitation,
where the angular momentum vector fluctuates
about the main rotation axis~\cite{ALL76}.
Now the reason of the small calculated $B(E2)$ ratio is clear
from the argument of the previous section:
The triaxiality $\gamma$ used in our calculations is
that of the Nilsson potential, $\gamma(\mbox{pot:Nils})$,
in \S\ref{sect:gamNils}, while the $\gamma$ used in the rotor model
is $\gamma(\mbox{den})$ in \S\ref{sect:gamQmom}.
As is discussed in the previous section, the difference between
them for the same shape is very large for large deformations
like in the case of the TSD bands; $\gamma(\mbox{pot:Nils})\approx 20^\circ$
corresponds to $\gamma(\mbox{den})\approx 11^\circ$,
and in order to perform the same calculation as the rotor model
with $\gamma(\mbox{den})\approx 20^\circ$ one has to employ
$\gamma(\mbox{pot:Nils})\approx 30^\circ$ according to Fig.~\ref{fig:gamNl}.
It should also be mentioned that we have used five major oscillator shells,
$N_{\rm osc}=3-5$ for proton and $N_{\rm osc}=4-8$ for neutron
in the calculation in Refs.~\cite{MSMwob1,MSMwob2,MSMwob3},
which were not enough; since the $i_{13/2}$ proton orbits are occupied
in the TSD band, the inclusion of $N_{\rm osc}=8$ proton quasiparticle
states are necessary in the RPA calculational step.

\begin{figure}[htbp]
\includegraphics[width=7cm,keepaspectratio]{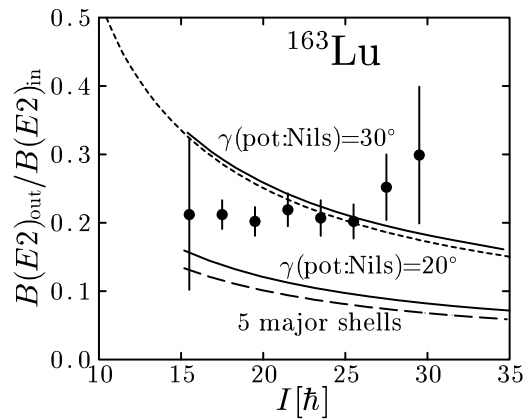}
\caption{
 The $B(E2)$ ratio,
 $B(E2\mbox{:}\,I\rightarrow I-1)_{\rm out}/
 B(E2\mbox{:}\,I\rightarrow I-2)_{\rm in}$,
 of the one-phonon wobbling band in $^{163}$Lu.
 The experimental data~\cite{wob163LuQ} is compared with
 the calculations by our microscopic RPA approach.
 The lower solid line is the result with
 $\gamma=\gamma(\mbox{pot:Nils})=20^\circ$, while the upper solid line
 is with $\gamma(\mbox{pot:Nils})=30^\circ$;
 the full model space is used for both of them.
 The dashed line is the result with $\gamma(\mbox{pot:Nils})=20^\circ$
 and using only the five major oscillator shells, corresponding to
 the previous calculation in Ref.~\cite{MSMwob1}.
 The dotted line is the same as that in Fig.~\ref{fig:rotBE2r},
 the particle-rotor calculation
 with $\gamma=\gamma(\mbox{den})=20^\circ$,
 depicted for a reference.
}
\label{fig:rpaBE2r}
\end{figure}

\begin{figure}[htbp]
\includegraphics[width=7cm,keepaspectratio]{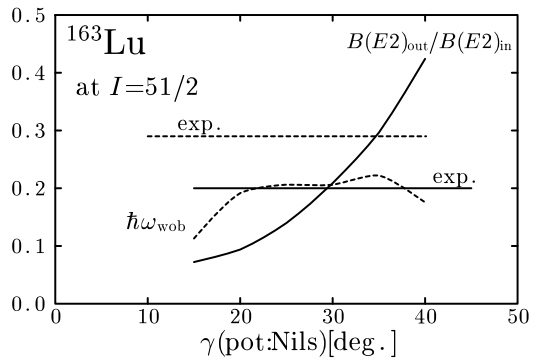}
\caption{
 The $\gamma$ dependence of the microscopically calculated
 $B(E2)$ ratio and the excitation energy
 of the one-phonon wobbling band in $^{163}$Lu at spin $I=51/2$.
 The solid lines are the $B(E2)$ ratios and the dotted lines
 are the energy in MeV.
 The horizontal solid and dotted lines designate the experimental values.
}
\label{fig:rpaOMB}
\end{figure}

In Fig.~\ref{fig:rpaBE2r} we depict the new results of calculation
employing $\gamma(\mbox{pot:Nils})=20^\circ$ and $30^\circ$
with using the full model space; all orbits in the oscillator
shell $N_{\rm osc}=0-9$ for both protons and neutrons are included.
The procedure and the other parameters in the calculation
are the same as in the previous work~\cite{MSMwob1};
$\epsilon_2=0.43$, $\epsilon_4=0$, and the pairing gaps
${\mit\Delta}_{\rm n,p}=0.3$ MeV.
The result of the previous calculation, the dashed line,
and that of the particle-rotor model with $\gamma(\mbox{den})=20^\circ$,
the dotted line, are also included.  Our previous calculation is smaller
than the experimental data partly because of the small model space,
but its effect is about 20\%; the large difference is mainly due to
the fact that we have used $\gamma=\gamma(\mbox{pot:Nils})=20^\circ$ in
the previous calculation, which corresponds to much smaller triaxiality
than $\gamma(\mbox{den})=20^\circ$ in the particle-rotor model.
The result with $\gamma(\mbox{pot:Nils})=30^\circ$ almost coincides
with that of the particle-rotor calculation
using $\gamma(\mbox{den})=20^\circ$, because the values
${\cal J}_x:{\cal J}_y:{\cal J}_z$ of the microscopically calculated
moments of inertia accidentally
take similar values in the relevant spin range~\cite{MSMwob1}.
In order to see the $\gamma$ dependence of the results,
we show, in Fig.~\ref{fig:rpaOMB},
the $B(E2)$ ratio and the excitation energy of
the one-phonon wobbling band in $^{163}$Lu at spin $I=51/2$
as functions of the triaxiality $\gamma=\gamma(\mbox{pot:Nils})$.
Although the excitation energy is rather flat in the range,
$20^\circ \le \gamma(\mbox{pot:Nils}) \le 30^\circ$,
the $B(E2)$ follows the behaviors of Eq.~(\ref{eq:BE2rot})
if the relation between $\gamma(\mbox{pot:Nils})$ and $\gamma(\mbox{den})$
is taken into account. The excitation energy can be expressed
in terms of the three moments of inertia in an usual way~\cite{BM75,Mar79},
but their $\gamma$ dependences are not simple like the irrotational
one~\cite{SM95} and lead to a rather weak $\gamma$ dependence
of the excitation energy in this case.
Here we have only shown the example of the wobbling excitation in $^{163}$Lu,
but we have confirmed that these properties of the wobbling-like RPA solution
are general, and can be applied for other cases in the Lu region.

Thus, our microscopic RPA calculations give more or less
the same results as that of the macroscopic particle-rotor model,
if the corresponding magnitude of the triaxial deformation is employed.
It shows that the RPA calculation of the wobbling excitation in the Lu region
leads to the behaviors of $B(E2)$'s that are given by the macroscopic
rotor model; namely the out-of-band $B(E2)$ can be related to
the static triaxial deformation.
This is non trivial since the out-of-band $B(E2)$ is calculated
by the RPA transition amplitudes of the non-diagonal part of
the quadrupole operators,
$Q_{21}^{(-)}$ and $Q_{22}^{(-)}$~\cite{Mar77,Mar79}.
It has been shown~\cite{SM95} that
the RPA wobbling theory of Marshalek~\cite{Mar79} gives
the same expression of the out-of-band $B(E2)$ as that of the rotor model,
if the RPA wobbling mode
is collective enough that the quantity ``$c_n$'' defined
in Eq.~(4.29) in Ref.~\cite{SM95} satisfies $c_{n={\rm wob}}=1$.
In the previous calculations~\cite{SM95,MSMwob1,MSMwob2,MSMwob3},
the employed model space was too small~\footnote{
 In Ref.~\cite{MSMwob3}, it was reported that $c_{n={\rm wob}}=0.6-0.8$,
 but these values were not correct;
 they were in the cases with even smaller model spaces.
 The calculation with the five major shells gives
 $c_{n={\rm wob}}\approx 0.9$, which leads to
 about a 20\% reduction as it is shown in Fig.~\ref{fig:rpaBE2r}
 of the present paper.
 }
to give $c_{n={\rm wob}}=1$,
but we have confirmed that $c_{n={\rm wob}}\approx 1$ is satisfied
within 1\% in the present full model space calculations.
Recently, this criterion, $c_{n={\rm wob}}\approx 1$, has been used
to identify the wobbling-like solution out of many RPA eigenmodes,
and shown to be very useful~\cite{KN07}.

\section{Summary}
\label{sect:sum}

In this work, we have first discussed the differences of
the various definitions of the triaxiality parameter $\gamma$.
The most basic among them is defined through
the two intrinsic quadrupole moments,
$\gamma(\mbox{den})$ in Eq.~(\ref{eq:gammaden})
for each configuration of a particular nucleus, or
$\gamma(\mbox{geo})$ in Eq.~(\ref{eq:gammageo})
for a given nuclear shape.
It has been found that these two coincide in a good approximation
(the nuclear shape consistency).
In the Hartree-Fock(-Bogoliubov) type calculations, where the nuclear
mean-field is determined selfconsistently
by a suitably chosen effective interaction,
the parameter $\gamma(\mbox{den})$ is the only possible
definition of triaxial deformation.
However, there is an another type of mean-field calculations, i.e.
the Strutinsky macroscopic-microscopic method, where one starts from
a suitably chosen average potential, in which the nuclear deformation
is parametrized in various different ways.
We have considered the two widely adopted ones,
the Nilsson type and the Woods-Saxon type parametrizations.
The triaxiality parameters associated with these potentials,
$\gamma(\mbox{pot:Nils})$ in Eqs.~(\ref{eq:Nilsdef})-(\ref{eq:Nilseps}) and
$\gamma(\mbox{pot:WS})$ in Eqs.~(\ref{eq:WSdef})-(\ref{eq:defRS}),
are compared with the density type
$\gamma(\mbox{den})$ and $\gamma(\mbox{geo})$.
Conspicuous differences between the potential type and the density type
$\gamma$'s, e.g. $\gamma(\mbox{pot:Nils})$ vs. $\gamma(\mbox{geo})$,
have been found especially for larger deformations, e.g.
the triaxial superdeformed states.
It is also investigated how the differences between various definitions
come out by evaluating their relations explicitly
in the small deformation limit.
Therefore we have to be very careful about which definition is used
in quantitative discussions of the triaxial deformation.

Next, we have investigated the out-of-band to in-band $B(E2)$ ratio
of the one-phonon wobbling band,
which is measured systematically in the Lu region and
is sensitive to the triaxial deformation.
The macroscopic particle-rotor model~\cite{HH03}
is used to deduce the triaxial deformation
from the experimental $B(E2)$ ratio,
which leads to $\gamma=\gamma(\mbox{den})\approx 20^\circ$ on average.
On the other hand, we performed the microscopic RPA
calculation~\cite{MSMwob1,MSMwob3}
with using $\gamma=\gamma(\mbox{pot:Nils})\approx 20^\circ$
corresponding to the TSD minima in the cranked Nilsson-Strutinsky
calculation~\cite{SP95,Bengt}, but we obtained
too small $B(E2)$ ratio compared with the experimental data.
It has been found that the reason of the underestimation of
our previous microscopic calculation is mainly due to the different triaxial
deformation used: We have used $\gamma(\mbox{pot:Nils})\approx 20^\circ$
corresponding roughly to $\gamma(\mbox{den})\approx 11^\circ$,
which is much smaller than $\gamma(\mbox{den})\approx 20^\circ$
in the rotor model calculations.
If the proper triaxiality corresponding to
$\gamma(\mbox{den})\approx 20^\circ$ is used,
our RPA calculation can nicely reproduce the magnitude of
the measured $B(E2)$ ratio in the same way as in the macroscopic
particle-rotor model.

It should, however, be emphasized that an important problem remains:
The predicted triaxial deformations, $\gamma(\mbox{den})\approx 11^\circ$,
by the cranked Nilsson-Strutinsky calculations~\cite{SP95,Bengt}
for the TSD bands in the Hf, Lu region are too small
to account for the measured $B(E2)$ ratio of the wobbling excitations.
We believe that this is a challenge to the existing microscopic theory.
An another thing we would like to mention is that the measured $B(E2)$,
both the out-of-band and in-band $B(E2)$'s, seems to indicate
that the triaxial deformation is changing as a function of spin;
it increases at higher spins~\cite{HH03}.
We have recently developed a new RPA approach~\cite{SS06}
based on the Woods-Saxon potential as a mean-field,
which is believed to be more reliable than our previous calculations
employing the Nilsson potential.  The result of calculations
and discussions including the issue of
the change of the triaxial deformation suggested by
the out-of-band as well as in-band $B(E2)$'s will be reported
in a subsequent paper; see Ref.~\cite{SS06} for a preliminary report.

\end{document}